# Four-dimensional image formation theory of optical coherence tomography

**NAOKI FUKUTAKE,**[1,2*] **SHUICHI MAKITA,**[2] **YOSHIAKI YASUNO** [2]

[1] *Nikon Corporation, 471 Nagaodai-cho, Sakae-ku, Yokohama-city, Kanagawa, 244–8533 Japan.*
[2] *University of Tsukuba, Tennodai 1–1-1, Tsukuba, Ibaraki, 305–8573 Japan.*
*\*Corresponding author: Naoki.Fukutake@nikon.com*

We construct an accurate imaging theory for optical coherence tomography/microscopy (OCT/OCM) without approximations to calculate precise optical resolution and imaging characteristics. Our theory represents a broadband light source using a four-dimensional (4D) pupil function with a dimension for light frequency (reciprocal of wavelength) as the fourth axis in 4D frequency space. Consequently, 4D space-time representation is required in real space, connected to the 4D frequency space by a 4D Fourier transform. Our theory provides insight into a peculiar image formation in OCT/OCM, particularly when an apparatus has a high numerical aperture (NA) optical system to handle aberrations and dispersions simultaneously.

## 1. Introduction

Optical coherence tomography/microscopy (OCT/OCM) has been widely used as a powerful bioimaging device and has established a firm position in the medical industry [1]. OCT has a 40-year history, with many types having been developed [2–5]. New ideas have been created through extensive experimentation. However, theories have not caught up with experiments as in other scientific fields. A precise theory is commonly formulated after diverse experimental results are obtained, which also holds true for OCT. Constructing a rigorous theory of OCT image formation is required to provide a correct understanding of the optical resolution and imaging characteristics.

Regarding the image formation theory of the OCT, which has a sample arm and a reference arm to use the interference between the light from the two arms, previous research has treated the optical resolution in the $z$ direction (optical axis direction) independent from that in the x-y directions [6]. Although some researchers have considered the mixing of x-y and z directions in an image of a single point object [7–9], they have equated the spatial variable $z$ with the delay time between the light from the two arms $\tau$ in their formulation, i.e., $z = c\tau$, where $c$ is the light speed. However, $z$ and $\tau$ are distinct variables and orthogonal to each other in the four-dimensional (4D) spatial-time coordinates.

In this paper, we construct a 4D theory that can rigorously treat the image formation of OCT/OCM. We define the 4D pupil functions of an excitation system and a collection system, in which one lens serves as the two systems. We also define a 4D aperture as an instrumental function determining the resolution limit calculated from the 4D pupil functions. Our formulation can treat time-domain OCT (TD-OCT) [1], Fourier-domain OCT (FD-OCT) [3,5,10,11], and full-field OCT (FF-OCT) [4,12]. Furthermore, our theory permits the simultaneous handling of aberration and dispersion. We will discuss how the 4D coordinate is used to examine the precise OCT image formation.

## 2. Image formation formula

We construct the image formation theory for some types of OCT/OCM in real space-time, where the 4D coordinate system $(x, y, z, c\tau) = (\mathbf{x}, c\tau)$ is applied. Although a three-dimensional (3D) molecular density distribution $N(x, y, z)$ is an object to be observed, the object itself cannot be obtained through a measurement. In our theory, the obtained data are referred to as an image (a 3D data set). In the image, some object information is permanently lost, causing a blurred image compared with the object. We demonstrate how the image data set is expressed using the 4D coordinate system, assuming the object is a transparent biomedical specimen without absorption.

**A. Point-scanning time-domain OCT**

Before expressing the 3D image data set, we consider a 4D meta-image $I(x', y', z', c\tau)$, where $(x', y', z') = \mathbf{x}'$ represents excitation-light spot displacement by a scanning mirror. Hereafter, we adopt the natural unit: $c = 1$. For convenience, we use the coordinates $(\mathbf{x}', \tau)$, but eventually we assume that $z' = 0$. Figure 1 (A) illustrates a schematic of the TD-OCT with coordinate systems, where $\mathbf{x} = (x, y, z)$ represents the 3D object coordinate, $t$ is the time when the signal light is emitted from a position $\mathbf{x}$ in the object, $t_\mathrm{d}$ is a detection time, and the delay $\tau$ corresponds to the difference in optical path length (OPL) between the reference and sample arms. The OPL of the sample arm is defined as the optical length from a beam waist of the excitation beam (geometrical focal plane), which is shifted to adjust the depth region to be observed, and we assume that a reference mirror in the reference arm is placed at the focal plane ($z = 0$).

In OCT, after the two beams from the sample and reference arms interfere, the square of the modulus of the interference amplitude, i.e., the intensity, is detected [1,13]. The meta-image intensity is expressed as

$$M_{TD}(\boldsymbol{x}', \tau) = \int \left| iL(t_d - \tau) + \int N(\boldsymbol{x} - \boldsymbol{x}')E_t(\boldsymbol{x}, t_d)d^3\boldsymbol{x} \right|^2 dt_d, \quad (1)$$

where $iL(t_d - \tau)$ is the complex light amplitude from the reference arm and the term that contains the $\boldsymbol{x}$ integral corresponds to the complex light amplitude from the sample arm. The phase $i = e^{i\pi/2}$ is derived from the reflection from the reference mirror. Because phase $i$ is not important in OCT, we omit it. $E_t(\boldsymbol{x}, t_d)$ represents an amplitude point spread function (aPSF) at the detection time $t_d$ for the total system that consists of an electric field distribution in the sample formed by the excitation system $E_{Se}(\boldsymbol{x}, t)$ and a dimensionless Green's function created by the signal-collection system $h_{Sc}(-\boldsymbol{x}, t_d - t)$:

$$E_t(\boldsymbol{x}, t_d) = \int E_{Se}(\boldsymbol{x}, t)h_{Sc}(-\boldsymbol{x}, t_d - t)dt. \quad (2)$$

The same lens serves as both the excitation and collection systems. An effective numerical aperture (NA) is determined by the width of Gaussian beam coming in/out from a single mode fiber and the NA cutoff of the lens system. Green's function $h_{Sc}(\boldsymbol{x}, t_d - t)$ corresponds to the complex probability amplitude that expresses a signal light propagation from a space-time point in the sample $(x, y, z, t)$ to a space-time point $(0, 0, 0, t_d)$ at the detection position, where a confocal system is assumed by the single-mode fiber. The electric field distribution of the reference beam $L(t_d - \tau)$, a simplified expression of $L(0, 0, 0, t_d - \tau)$, is also influenced by the confocality. Although the light-spot scan is usually applied in OCT, we assume a sample-stage scan, as depicted in Eq. 1 (A), which generates the same meta-image as the spot-scanning OCT system.

In OCT, because the $z'$ scan is not applied ($z' = 0$), the 4D meta-image $I(x', y', z', \tau)$ reduces to the 3D image $I_{TD}(x', y', \tau)$ after extracting one of the cross-terms in Eq. (1):

$$I_{TD}(x', y', \tau)$$
$$= \int L^*(t_d - \tau)\left\{\int N(\boldsymbol{x} - \boldsymbol{x}')E_t(\boldsymbol{x}, t_d)d^3\boldsymbol{x}\right\}dt_d \Big|_{z'=0}. \quad (3)$$

It can be assumed that the detection time at each scanning point $(x', y', \tau)$ is sufficiently long compared with the light frequency. In The integral over $t_d$ in Eq. (3) becomes the cross-correlation between $L^*(t_d - \tau)$ and $E_t(\boldsymbol{x}, t_d)$. After the calculation of the cross-correlation, we obtain

$$I_{TD}(x', y', \tau) = \int N(\boldsymbol{x} - \boldsymbol{x}')PSF_4(\boldsymbol{x}, \tau)d^3\boldsymbol{x} \Big|_{z'=0}, \quad (4)$$

where

$$PSF_4(\boldsymbol{x}, \tau) = \int L^*(t_d - \tau)E_t(\boldsymbol{x}, t_d)dt_d \quad (5)$$

includes the information on the 3D intensity point spread functions $PSF_3(x, y, \tau: z_0)$ with different shapes depending on the depth $z_0$ in the sample. For instance, if the object is a single point on the optical axis $N(\boldsymbol{x} - \boldsymbol{x}') = \delta(x - x')\delta(y - y')\delta(z - z_0 - z')$, Eq. (4) becomes $I_{TD}(x', y', \tau) = PSF_4(x', y', z_0, \tau)$, which corresponds to $PSF_3(x, y, \tau: z_0)$. This is a 3D function of $(x', y', \tau)$ that changes the shape along with the depth $z_0$ as depicted later.

We consider the reference beam in more detail. The complex amplitude of the beam propagating through the reference arm onto the single-mode fiber can be written as

$$L(t_d - \tau) = \int E_{Re}(\boldsymbol{x}_2, t_2)\delta(z_2)$$
$$\times h_{Rc}(\boldsymbol{x}_d - \boldsymbol{x}_2, t_d - t_2 - \tau)d^3\boldsymbol{x}_2 dt_2 \Big|_{\boldsymbol{x}_d=0}, \quad (6)$$

where $E_{Re}(\boldsymbol{x}_2, t_2)$ and $h_{Rc}(\boldsymbol{x}_d - \boldsymbol{x}_2, t_d - t_2 - \tau)$ are the excitation electric field in the reference arm and Green's function for the reference beam collection. $\boldsymbol{x}_d = \boldsymbol{0} = (0,0,0)$ is the position of the single-mode fiber, and $\delta(z_2)$ indicates that the reference mirror is placed at the focal plane of the lens ($z_2 = 0$). The delay time $\tau$ is not a variable in the delta function $\delta(z_2)$ but in Green's function $h_{Rc}(\boldsymbol{x}_d - \boldsymbol{x}_2, t_d - t_2 - \tau)$ because the lens and the reference mirror are scanned together as a unit [Fig. 1 (A)].

### B. Point-scanning frequency-domain OCT

Figure 1 (B) illustrates a schematic of the FD-OCT with coordinate systems. In FD-OCT, while the reference mirror is fixed, the light frequency of a light source is scanned [5,11], or the interference light between the two beams from the sample and reference arms is measured by a spectrometer [3,10]. In both cases, the meta-image intensity at each light frequency $\nu$ can be expressed as

$$I_\nu(\boldsymbol{x}', \nu) = \left| i\tilde{L}(\nu) + \int N(\boldsymbol{x} - \boldsymbol{x}')E_\nu(\boldsymbol{x}, \nu)d^3\boldsymbol{x} \right|^2, \quad (7)$$

where $z'$ eventually sets as zero, $\tilde{L}(\nu)$ represents a complex amplitude spectrum of the reference beam corresponding to the Fourier transform of $L(t_d)$, and $E_\nu(\boldsymbol{x}, \nu)$ is the aPSF for the total system that is generated by a light frequency $\nu$, which has a Fourier transform relationship with $E_t(\boldsymbol{x}, t_d)$ as follows:

$$E_\nu(\boldsymbol{x}, \nu) = \int E_t(\boldsymbol{x}, t_d) e^{i2\pi\nu t_d}dt_d. \quad (8)$$

In considering the FD-OCT image, the measured intensity at each light frequency $I_\nu(\boldsymbol{x}', \nu)$ is Fourier transformed:

$$M_{FD}(\boldsymbol{x}', \tau) = \int I_\nu(\boldsymbol{x}', \nu)e^{-i2\pi\nu\tau}d\nu, \quad (9)$$

which corresponds to the meta-image of FD-OCT. Then, a cross-term of the meta-image $M_{FD}(\boldsymbol{x}', \tau)$ is extracted, expressed as:

$$I_{FD}(x', y', \tau)$$
$$= \int \tilde{L}^*(\nu)\left\{\int N(\boldsymbol{x} - \boldsymbol{x}')E_\nu(\boldsymbol{x}, \nu)d^3\boldsymbol{x}\right\}e^{-i2\pi\nu\tau}d\nu \Big|_{z'=0}$$
$$= \int N(\boldsymbol{x} - \boldsymbol{x}')PSF_4(\boldsymbol{x}, \tau)d^3\boldsymbol{x} \Big|_{z'=0}, \quad (10)$$

with

$$PSF_4(\boldsymbol{x}, \tau) = \int \tilde{L}^*(\nu) E_\nu(\boldsymbol{x}, \nu)e^{-i2\pi\nu\tau}d\nu, \quad (11)$$

where $z'$ is set to zero, and $\tau$ is a variable defined in Fourier transform ($\nu \leftrightarrow \tau$) on a computer. This formula for FD-OCT becomes equal to that for TD-OCT expressed in Eq. (4). In the experimental setup of FD-OCT, a physical delay time offset $\Delta\tau$ is adjusted to separate the two cross-terms of Eq. (7).

### C. Kohler-illumination time-domain OCT

We then represent FF-OCT [4,12], where Kohler illumination [14] is applied in the time domain, as depicted in Fig. 1 (C). In FF-OCT, instead of sample-stage or light-spot scanning, a two-dimensional (2D) detector is used, which indicates that the coordinate $(x', y')$ corresponds to a position on the 2D detector. The meta-image intensity can be expressed as

$$M_{FF}(\boldsymbol{x}', \tau) = \int |iR(\boldsymbol{x}', t_d - \tau) + S(\boldsymbol{x}', t_d)|^2 dt_d, \quad (12)$$

where

$$R(\pmb{x}', t_{\mathrm{d}} - \tau) = \int K_{\mathrm{R}}(\pmb{x}_2, t_2)\, \delta(z_2)$$
$$\times h_{\mathrm{Rc}}(\pmb{x}' - \pmb{x}_2, t_{\mathrm{d}} - t_2 - \tau)\, d^3\pmb{x}_2 dt_2 \quad (13)$$

represents the complex light amplitude from the reference arm,

$$S(\pmb{x}', t_{\mathrm{d}}) = \int K_{\mathrm{S}}(\pmb{x}_1, t_1) N(\pmb{x}_1)$$
$$\times h_{\mathrm{Sc}}(\pmb{x}' - \pmb{x}_1, t_{\mathrm{d}} - t_1)\, d^3\pmb{x}_1 dt_1 \quad (14)$$

is the complex light amplitude given the contribution from the sample arm, $h_{\mathrm{Rc}}(\pmb{x}' - \pmb{x}_2, t_{\mathrm{d}} - t_2 - \tau)$ and $h_{\mathrm{Sc}}(\pmb{x}' - \pmb{x}_1, t_{\mathrm{d}} - t_1)$ correspond to Green's function of the collection systems in the sample and reference arms, and $K_{\mathrm{S}}(\pmb{x}_1, t_1)$ and $K_{\mathrm{R}}(\pmb{x}_2, t_2)\delta(z_2)$ indicate the complex light amplitude in the sample and the complex amplitude on the reference mirror shined by the Kohler-illumination system.

One of the cross-terms in Eq. (12) $\int R^*(\pmb{x}', t_{\mathrm{d}} - \tau) S(\pmb{x}', t_{\mathrm{d}}) dt_{\mathrm{d}}$ provides the 4D meta-image of FF-OCT. Before calculating the cross-term, we illustrate the relationship between the light amplitude by the Kohler illumination $K$ and the complex coherence function defined in classical optics $\Gamma$ [15]:

$$K_{\mathrm{S}}(\pmb{x}_1, t_1) K_{\mathrm{R}}^*(\pmb{x}_2, t_2) = \Gamma(\pmb{x}_1 - \pmb{x}_2, t_1 - t_2)$$
$$= \Gamma^*(\pmb{x}_2 - \pmb{x}_1, t_2 - t_1). \quad (15)$$

We then define the effective excitation point spread function (PSF) as

$$I_{\mathrm{eff}}^*(\pmb{x}' - \pmb{x}_1, t_d - t_1 - \tau)$$
$$= \int \Gamma^*(\pmb{x}_2 - \pmb{x}_1, t_2 - t_1) \delta(z_2)$$
$$\times h_{\mathrm{Rc}}^*(\pmb{x}' - \pmb{x}_2, t_d - t_2 - \tau)\, d^3\pmb{x}_2 dt_2, \quad (16)$$

which leads to the instrumental function in the total system of FF-OCT as follows:

$$PSF_{\mathrm{T}4}(\pmb{x}' - \pmb{x}, \tau) = \int I_{\mathrm{eff}}^*(\pmb{x}' - \pmb{x}, T - \tau)$$
$$\times h_{\mathrm{Sc}}(\pmb{x}' - \pmb{x}, T) dT, \quad (17)$$

where the variable conversion $T = t_d - t_1$ is applied. Finally, we obtain the cross-terms $C_{\mathrm{FF}}(\pmb{x}', \tau)$ in the 4D meta-image as follows:

$$C_{\mathrm{FF}}(\pmb{x}', \tau) = \int N(\pmb{x}) PSF_{\mathrm{T}4}(\pmb{x}' - \pmb{x}, \tau) d^3\pmb{x} dt_{\mathrm{d}}$$
$$= \Delta t_{\mathrm{d}} \int N(\pmb{x}' - \pmb{x}) PSF_{\mathrm{T}4}(\pmb{x}, \tau) d^3\pmb{x}, \quad (18)$$

where $\Delta t_{\mathrm{d}}$ is the exposure time, which is sufficiently long compared with the period of the light. In formulating the FF-OCT image formation in the same dimension as TD- and FD-OCT, a relationship of

$$PSF_4(-\pmb{x}, \tau) = \Delta t_{\mathrm{d}} PSF_{\mathrm{T}4}(\pmb{x}, \tau) \quad (19)$$

can be used. We derive the image formation formula of FF-OCT:

$$I_{\mathrm{FF}}(x', y', \tau) = \int N(\pmb{x} - \pmb{x}') PSF_4(\pmb{x}, \tau) d^3\pmb{x} \bigg|_{z'=0}. \quad (20)$$

The space coordinates in $PSF_4(-\pmb{x}, \tau)$ are reversed because of the difference between the sample-stage scanning system and the 2D detector system with the sample fixed. For convenience, $PSF_4(\pmb{x}, \tau)$ is separated into two parts in the same manner as other OCT types as follows:

$$PSF_4(\pmb{x}, \tau) = \int L^*(t_{\mathrm{D}} - \tau) E_t(\pmb{x}, t_{\mathrm{D}}) dt_{\mathrm{D}}, \quad (21)$$

where $E_t(\pmb{x}, t)$ and $L(t)$ have the same dimension as the electric field.

### D. General formula

Equations (4), (10), and (20) exhibit the same formula, indicating that TD-, FD-, and FF-OCT generate the same image if the experimental conditions are adjusted. Therefore, we can uniformly discuss OCT image formation. Irrespective of the type of OCT, a 4D point spread function (4D PSF) is defined as $PSF_4(\pmb{x}, \tau)$. With TD-OCT as a representative example, the 4D PSF can be expressed as

$$PSF_4(\pmb{x}, \tau) = \int E_{\mathrm{Se}}(\pmb{x}, t) E_{\mathrm{c}}(-\pmb{x}, \tau - t) dt, \quad (22)$$

where

$$E_{\mathrm{c}}(\pmb{x}, \tau - t) = \int L^*(t_{\mathrm{d}} - \tau) h_{\mathrm{Sc}}(-\pmb{x}, t_d - t) dt_{\mathrm{d}} \quad (23)$$

has the dimensions of the electric field, and its bandwidth is restricted by the reference light frequency. The dimensionless Green's function $h_{\mathrm{Sc}}(-\pmb{x}, t_d - t)$ accepts all light frequencies. The 4D PSF can be expressed as the product of two beams: the excitation and collection beams [see Eq. (22)] with respect to space and convolution with respect to time. The collection beam direction, which is opposite to the excitation beam, $(-\pmb{x}, \tau - t)$ propagates in the same direction as $E_{\mathrm{ex}}(\pmb{x}, t)$ because of $-\pmb{x}$ in the collection beam. Therefore, the complex amplitude of the 4D PSF becomes equal to the square of the complex amplitude of a focusing beam under the condition of the same effective NA for both the excitation and collection systems. The 4D PSF provides the characteristics of the image formation, as depicted in Fig. 2.

## 3. Four-dimensional frequency space

We intuitively interpret the image formation using 4D frequency space, which consists of three spatial-frequency axes $\pmb{f} = (f_x, f_y, f_z)$ and one light-frequency axis $\nu$. The 4D frequency space $(\pmb{f}, \nu)$ is connected to real space-time $(\pmb{x}, t)$ by a 4D Fourier transform:

$$\tilde{g}(\pmb{f}, \nu) = \int g(\pmb{x}, t) e^{i2\pi(f_x x + f_y y + f_z z - \nu t)}\, d^3\pmb{x} dt. \quad (24)$$

In OCT, the fourth argument relating to time is the delay time $\tau$, i.e., $t \to \tau$.

### A. Four-dimensional pupil function

We begin with defining 4D pupil functions in the frequency space. We expand the concept of the conventional 3D pupil function [16] to the 4D pupil function by adding the light frequency ($\nu$) axis, as illustrated in Fig. 3 (A). In OCT, because the first order light-matter interaction [15], i.e., reflection, is used, two pupils exist in each arm: a sample-excitation pupil and a signal-collection pupil in the sample arm, and a reference excitation pupil and a reference-collection pupil in the reference arm.

In OCT, the same optical system serves as both the excitation and collection systems in both the sample and reference arms. Furthermore, optical systems with identical specifications are used for both arms. Consequently, the four pupils possess the same effective NA. Although the light frequency ($\nu$) direction of the excitation pupils in both arms is restricted by the light amplitude spectrum of the light source, the collection pupils in both arms work for all light frequencies. They are shaped like a part of a conic surface restricted only by the effective NA in the 4D frequency space, as depicted in Fig. 3 (A).

In TD- and FD-OCT, the 4D sample-excitation pupil function $P_{\mathrm{Se}}(\pmb{f}, \nu)$ is calculated by Fourier transforming the excitation beam $E_{\mathrm{Se}}(\pmb{x}, t)$. In FF-OCT, the effective excitation pupil function $P_{\mathrm{eff}}(\pmb{f}, \nu)$ can be defined as the 4D Fourier transform of the excitation-light amplitude function that removes the contribution of $L^*(\tau)$ from the effective excitation PSF, $I_{\mathrm{eff}}^*(\pmb{x}, \tau)$. Thus, $\tilde{L}^*(\nu) P_{\mathrm{eff}}(\pmb{f}, \nu)$ corresponds to the Fourier transform of $I_{\mathrm{eff}}^*(\pmb{x}, \tau)$, where $\tilde{L}^*(\nu)$ corresponds to the amplitude spectrum from a single point on the light source. It is notable that the excitation pupil function in TD- and FD-OCT, $P_{Se}(\boldsymbol{f},\nu)$, and the effective excitation pupil function in FF-OCT, $P_{eff}(\boldsymbol{f},\nu)$, become identical if the NA and the amplitude spectrum of the light source are adjusted.

Before defining the effective excitation pupil function $P_{eff}(\boldsymbol{f},\nu)$ in FF-OCT, we must consider a Kohler-illumination pupil function from a 2D incoherent light source. The light amplitude $K(\boldsymbol{x},t)$ is expressed as the inverse Fourier transform of the Kohler-illumination pupil function $P_{ill}(\boldsymbol{f},\nu)$, which has random phases because of the 2D incoherent light source:

$$K(\boldsymbol{x},t) = \int P_{ill}(\boldsymbol{f},\nu)e^{i2\pi(\boldsymbol{f}\cdot\boldsymbol{x}-\nu t)}d^3\boldsymbol{f}d\nu. \quad (25)$$

Given this relationship, the complex coherence function can be expressed as

$$\begin{aligned}\Gamma(\boldsymbol{x}_1-\boldsymbol{x}_2,t_1-t_2) \\ = \int P_{ill}^{(S)}(\boldsymbol{f}_1,\nu_1)e^{i2\pi(\boldsymbol{f}_1\cdot\boldsymbol{x}_1-\nu_1 t_1)}d^3\boldsymbol{f}_1 d\nu_1 \\ \times \int \{P_{ill}^{(R)}(\boldsymbol{f}_2,\nu_2)\}^* e^{-i2\pi(\boldsymbol{f}_2\cdot\boldsymbol{x}_2-\nu_2 t_2)}d^3\boldsymbol{f}_2 d\nu_2,\end{aligned} \quad (26)$$

where $P_{ill}^{(S)}(\boldsymbol{f}_1,\nu_1)$ and $P_{ill}^{(R)}(\boldsymbol{f}_2,\nu_2)$ correspond to the Fourier transform of $K_S(\boldsymbol{x}_1,t_1)$ and $K_R(\boldsymbol{x}_2,t_2)$, i.e., the Kohler-illumination pupil functions in the sample and reference arms. Under usual experimental conditions, the two Kohler-illumination pupil functions are identical to $P_{ill}(\boldsymbol{f},\nu)$:

$$\begin{aligned}P_{ill}^{(S)}(\boldsymbol{f}_1,\nu_1)\{P_{ill}^{(R)}(\boldsymbol{f}_2,\nu_2)\}^* \\ = \delta^3(\boldsymbol{f}_1-\boldsymbol{f}_2)\delta(\nu_1-\nu_2)|P_{ill}(\boldsymbol{f},\nu)|^2.\end{aligned} \quad (27)$$

Accordingly, the complex coherence function can be expressed as

$$\Gamma(\boldsymbol{x}_1-\boldsymbol{x}_2,t_1-t_2) = \int |P_{ill}(\boldsymbol{f},\nu)|^2 e^{i2\pi\{\boldsymbol{f}\cdot(\boldsymbol{x}_1-\boldsymbol{x}_2)-\nu(t_1-t_2)\}}d^3\boldsymbol{f}d\nu. \quad (28)$$

We then calculate the Fourier transform of the effective excitation PSF:

$$\begin{aligned}P_{Iex}(\boldsymbol{f},\nu) &= \int I_{eff}(\boldsymbol{x},t)e^{-i2\pi(\boldsymbol{f}\cdot\boldsymbol{x}-\nu t)}d^3\boldsymbol{x}dt \\ &= \{\tilde{\Gamma}(f_x,f_y,f_z,\nu)\otimes b(f_x,f_y,\nu)\}\{P_{Rc}(\boldsymbol{f},\nu)\}^*,\end{aligned} \quad (29)$$

where $\tilde{\Gamma}(f_x,f_y,f_z,\nu)$ corresponds to an effective light source [17], $b(f_x,f_y,\nu) = \delta(f_x)\delta(f_y)\delta(\nu)$ is a straight line function in the $f_z$ direction, $P_{Rc}(\boldsymbol{f},\nu)$ represents the reference-collection pupil function, and $\otimes$ is the 4D convolution. Figure 3 (B) illustrates the calculation process of $P_{Iex}(\boldsymbol{f},\nu)$. Finally, we obtain the effective excitation pupil function $P_{eff}(\boldsymbol{f},\nu)$ in FF-OCT as follows:

$$P_{Iex}(\boldsymbol{f},\nu) = \tilde{L}^*(\nu)P_{eff}(\boldsymbol{f},\nu), \quad (30)$$

where we assume that $P_{eff}(\boldsymbol{f},\nu)$, $\tilde{L}^*(\nu)$, and $P_{Iex}(\boldsymbol{f},\nu)$ have the dimensions of the electric field, the complex conjugate to the electric field, and the intensity, respectively. The effective excitation pupil is moved to the opposite side with respect to the $f_z$ axis.

In TD- and FD-OCT, the pupil functions for the reference arm can also be considered similarly. By Fourier transforming Eq. (6), the amplitude spectrum of the reference beam can be obtained as follows:

$$\tilde{L}(\nu) = \int \{P_{Re}(\boldsymbol{f},\nu)\otimes b(f_x,f_y,\nu)\}\{P_{Rc}(\boldsymbol{f},\nu)\}d^3\boldsymbol{f}, \quad (31)$$

where $P_{Re}(\boldsymbol{f},\nu)$ and $P_{Rc}(\boldsymbol{f},\nu)$ are the 4D excitation and collection pupil functions in the reference arm.

The signal-collection pupil function in the sample arm $P_{Sc}(\boldsymbol{f},\nu)$ is standard for all types of OCT.

### B. Four-dimensional aperture
We define the Fourier components that contribute to image formation as a 4D aperture $A_4(\boldsymbol{f},\nu)$ in the 4D frequency space. In the 4D frequency space, we consider the object frequency (Fourier components) of the molecular density distribution $N(\boldsymbol{x})$ that is uniform in the $\nu$-axis direction. Only the Fourier components of the object inside the 4D aperture can be captured by the OCT system, and the Fourier components form an image. After Fourier transforming Eqs. (4), (10), and (20), an image frequency $\tilde{I}(f_x,f_y,\nu)$ can be calculated as follows:

$$\tilde{I}(f_x,f_y,\nu) = \int \tilde{N}(\boldsymbol{f})A_4(\boldsymbol{f},\nu)df_z, \quad (32)$$

where $\tilde{N}(\boldsymbol{f})$ represents the object frequency, the 4D aperture $A_4(\boldsymbol{f},\nu)$ is calculated by 4D Fourier transforming the 4D PSF $PSF_4(\boldsymbol{x},\tau)$, and the integration over $f_z$ results from the $z'$ scan not being applied ($z' = 0$). Thus, the 4D aperture [Fig. 3 (C)] becomes an indicator for the resolution limit, providing a frequency cutoff. The positive direction of $\boldsymbol{f}$ is redefined according to $\tilde{N}(\boldsymbol{f})$ because the axis direction is flipped after the Fourier transform [see Eqs. (4), (10), and (20)].

In calculating the 4D aperture, we simultaneously consider the Fourier transform of $PSF_4(\boldsymbol{x},\tau)$ for TD-, FD-, and FF-OCT, expressed in Eqs. (5), (11), and (21). All 4D apertures are calculated similarly using two pupil functions, i.e., the signal-collection pupil $P_{Sc}(\boldsymbol{f},\nu)$ and the excitation pupil $P_e(\boldsymbol{f},\nu)$. For FF-OCT, the effective excitation pupil $P_{eff}(\boldsymbol{f},\nu)$ is used as the excitation pupil, while for TD- and FD-OCT, $P_{Se}(\boldsymbol{f},\nu)$ is used. For the 4D aperture calculation, we perform the convolution of the signal-collection pupil and the excitation pupil $P_e(-\boldsymbol{f},\nu)\otimes P_{Sc}(\boldsymbol{f},\nu)$ in three dimensions $(f_x,f_y,f_z)$ at each light frequency $\nu$, deriving from Eqs. (2) and (17) and considering $\boldsymbol{f}$-axes flip mentioned above. In a physical sense, the signal emission time $t$ is uncertain and produces an unchanging signal wavelength through the propagation in the signal-collection system. We then multiply the reference frequency $\tilde{L}^*(\nu)$ by $P_e(-\boldsymbol{f},\nu)\otimes P_{Sc}(\boldsymbol{f},\nu)$ to obtain the 4D aperture, as depicted in Fig. 3 (C).

### C. Image formation in the frequency-space representation
The object frequency captured by a 4D aperture is defined as a 4D mate image frequency. We consider 3D image formation in the 4D frequency space. According to Eq. (32), the 4D mate image frequency is integrated over $f_z$ to form a 3D image, as depicted in Fig. 3 (D). In the frequency space, while the object frequency is a function of $(f_x,f_y,f_z)$, the image frequency is a function of $(f_x,f_y,\nu)$. Consequently, the object in the real space is a function of $(x,y,z)$, while the image is a function of $(x',y',\tau)$. A 3D Fourier transform between the functions of $(f_x,f_y,\nu)$ and $(x',y',\tau)$ is performed.

## 4. Discussions
The 4D PSF indicates that the image of a single point object outside a depth of focus (DOF) becomes a curved shape if the effective NA is sufficiently large and the light-source spectrum is sufficiently wide, as depicted in Fig. 2. Consequently, the coherence gate is curved in the image space $(x',y',\tau)$. In a low effective NA system, the curved shape of PSF is not outstanding, and thus, this effect can be ignored.

In addition to the distorted image, the high-NA system may lose some object-structure information. Although some digital refocusing techniques have been developed [9,18], the loss of object-structure information causes the imperfection of refocusing. Consequently, we consider the 4D meta-image frequency $\widetilde{N}(\boldsymbol{f})A_4(\boldsymbol{f},v)$ in the 4D frequency space [Fig. 3 (D)]. When the 3D image frequency is formed by the OCT system, the 4D meta-image frequency is integrated over $f_z$. If the NA is sufficiently large, the 4D meta-image frequency has a large thickness in the $f_z$ direction, indicating that a part of the object frequency is added together, becoming inseparable. If the effective excitation NA in FF-OCT is almost zero, i.e., a plane wave excitation, the 4D aperture becomes sufficiently thin in the $f_z$ direction, which implies a perfect refocusing through a double projection from $(f_x, f_y, v)$ space to $(f_x, f_y, f_z)$ space via the thin 4D aperture [Fig. 4]. After perfect refocusing, the distorted image of a single point outside the DOF returns to the same image as that of a focal point, i.e., the equivalent image to the microscope image with $z'$ scan.

The aberration and dispersion can simultaneously be considered a 4D aberration using the 4D pupil, where we define the conventional monochromatic aberration as a spatial aberration and the conventional dispersion as a time aberration. Two pupils out of the four pupils, i.e., the sample-excitation, signal-collection, reference-illumination, and reference-collection pupils, influence the spatial aberration of the total system, in which the combination of the two pupils differs depending on the OCT type, as presented in Table 1. In any type of OCT, the time aberration is canceled out if the dispersions of the sample and reference arms are adjusted to be identical.

## 5. Conclusions

We have developed an OCT image formation theory to handle TD-OCT, FD-OCT, and partially coherent FF-OCT. We demonstrated the necessity of a 4D formulation to examine precise OCT image formation characteristics. In the 4D frequency space, we defined the 4D pupil function as a part of the conic surface and the 4D aperture as a window function capturing the object Fourier components contributing to the image formation. With the 4D frequency representation, the OCT image formation can be considered the $f_z$ integration of the object frequency captured by the 4D aperture.

**Acknowledgment**
This work was supported by Nikon Corporation (Fukutake), Core Research for Evolutional Science and Technology (JPMJCR2105, Yasuno), and Japan Society for the Promotion of Science (21H01836 (Yasuno), 22K04962 (Makita)).

**Disclosures**
The authors declare no conflicts of interest.

**Competing interests**
The authors declare no competing interests. The authors (N. Fukutake, Y. Yasuno, S. Makita) published a Japanese patent (Publication #: WO2023/182011) and applied for a US patent (Application #: US 18/892126) related to a part of OCT.

**Figures and table**

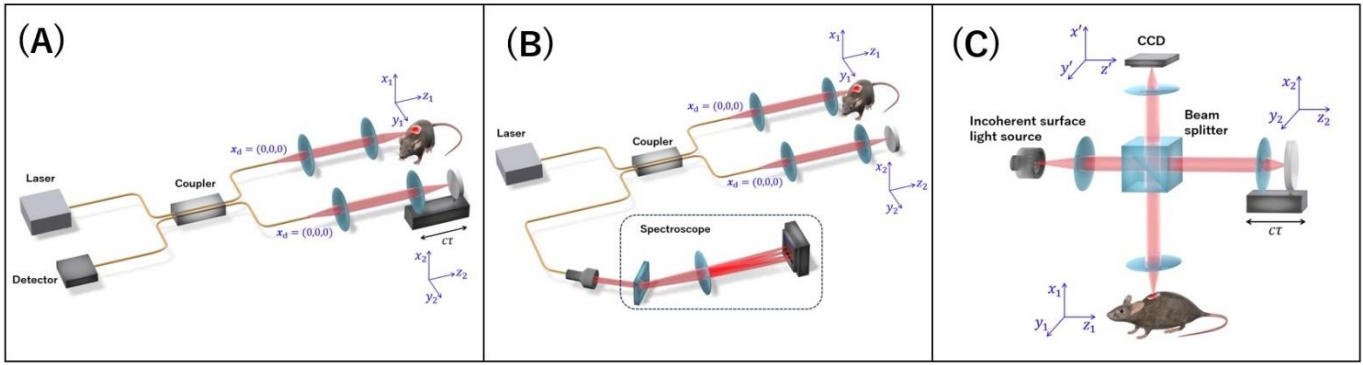

Fig. 1. Schematic of OCT. (A) Time-domain OCT. The light-spot scanning mirror is omitted. (B) Fourier-domain OCT. The light-spot scanning mirror is omitted. (C) Full-field OCT.

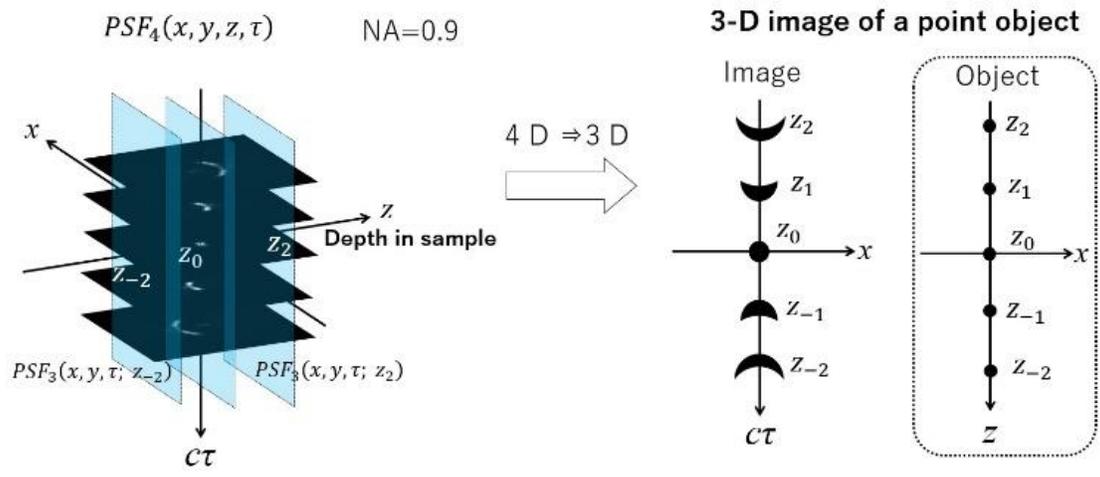

Fig. 2. Illustration of 4D PSF. The 4D PSF reduces to 3D PSF in OCT.

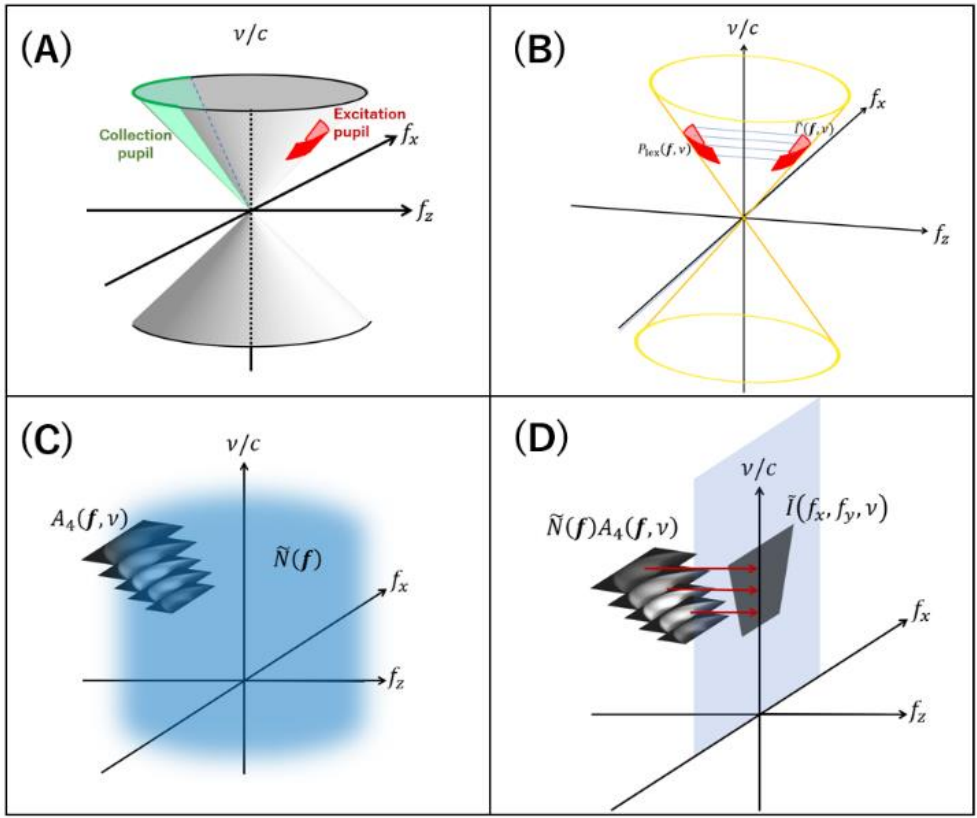

Fig. 3. 4D frequency-space representation. (A) 4D pupil function. (B) Illustration for the calculation process of $P_{lex}(f, v)$ in FF-OCT. (C) 4D aperture. (D) OCT image formation in the 4D frequency space.

Table 1. Contribution of pupil functions to aberration.

|  | Sample excitation | Sample collection | Reference excitation | Reference collection |
|---|---|---|---|---|
| **TD-OCT** | contribute | contribute |  |  |
| **FD-OCT** | contribute | contribute |  |  |
| **FF-OCT** |  | contribute |  | contribute |

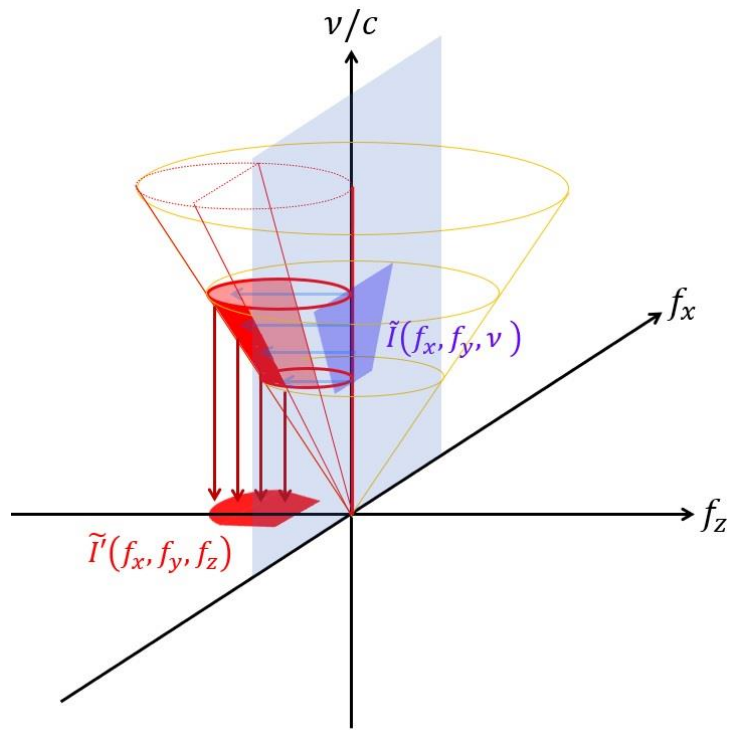

Fig. 4. Perfect refocusing with a thin 4D aperture. This condition can be satisfied by zero NA excitation.